# Stéphane Leduc and the vital exception in the Life Sciences

Raphaël Clément *

Institut de Biologie du Développement de Marseille

Aix-Marseille Université / CNRS UMR7288

13009 Marseille, France

**Abstract**

**Embryogenesis, the process by which an organism forms and develops, has long been and still is a major field of investigation in the natural sciences. By which means, which forces, are embryonic cells and tissues assembled, deformed, and eventually organized into an animal? Because embryogenesis deeply questions our understanding of the mechanisms of life, it has motivated many scientific theories and philosophies over the course of history. While genetics now seems to have emerged as a natural background to study embryogenesis, it was intuited long ago that it should also rely on mechanical forces and on the physical properties of cells and tissues. In the early 20$^{th}$ century, Stéphane Leduc proposed that biology was merely a subset of fluid physics, and argued that biology should focus on how forces act on living matter. Rejecting vitalism and life-specific approaches, he designed naïve experiments based on osmosis and diffusion to mimic shapes and phenomena found in living systems, in order to identify physical mechanisms that could support the development of the embryo. While Leduc's ideas then had some impact in the field, notably on later acclaimed D'Arcy Thompson, they fell into oblivion during the later 20$^{th}$ century. In this article I give an overview of Stéphane Leduc's physical approach to life, and show that the paradigm that he introduced, although long forsaken, becomes more and more topical today, as developmental biology increasingly turns to physics and self-organization theories to study the mechanisms of embryogenesis. His story, I suggest, bears witness to our reluctance to abandon life-specific approaches in biology.**

**Introduction**

Understanding how living organisms develop and organize during the embryonic period has always been, and remains, a major challenge of the natural sciences. Two main paradigms have long opposed to explain the development of the embryo. Epigenesis refers to the process by which organisms develop from a seed or an egg, and assumes that over the course of development, the embryo progressively acquires new parts, forms new organs and complexifies, to eventually become a *formed* animal. The theory of preformation, on the opposite, assumed that organisms grow by simple deployment of already pre-existing structures, and thus develop from a miniature version of themselves. Epigenesis was the prevalent theory until the 17$^{th}$ century, and the first observations of spermatozoa by Leeuwenhoek **(Leeuwenhoek 1677)**. After that the epigenesis became harder to defend: how could complex animals develop from structures as simple as sperm cells? Rather, it was proposed that preformed miniature animals should be already present in the reproductive cells, either female (ovism) or male (animalculism). From the 17$^{th}$ to 19$^{th}$ century, the theory of preformation would thus prevail over epigenesis. Importantly, preformation was also supported by the Christian Church: as the creator of all things, God created *all* men (and animals), including their miniature descendants inside them, themselves carrying their descendants, etc. Diderot and D'Alembert's Encyclopedia gives a flavor of the state of the art in the mid-18$^{th}$ century. Under *Génération (Physiologie)*, the authors present ovism and animalculism as the two main theories of embryogenesis **(D'Alembert, Mallet and d'Aumont 1757)**. "The two theories of generation […], the theory of eggs, as containing rudiments of the fetus, and that of spermatic worms, as forming these rudiments themselves, have divided almost all physicists since about a century". Still, later in the article they mention *correspondence:* raphael.clement@univ-amu.fr

epigenetic theories, such as that of Maupertuis, based on attraction: "A few modern physicists sought to find in the opinions of ancients more satisfying explanations about the mystery of generation. The author of the *Venus Physique* proposed to come back to the mixture of the two semens, that from the woman, and that from the man; and to account for the result of this mixture, he relies on attraction: why, he says, if this force exists in nature, should it be absent from the generation of animals? If there are in each of these semens parts meant to form the heart, the gut, the head, the arms, the legs, and that these parts each have a greater attraction to the part which should be its neighbor in the formation of the animal, the fetus will form." Having described the rich panel of theories of generation, the authors conclude: "we do not need more to prove that the mystery on this subject is impenetrable in nature." About fifty years later, at the beginning of the 19th century, Heinz Christian Pander, a German embryologist studying the chick embryo, discovered the existence of germ layers, distinct regions of the early embryo that later give rise to distinct tissues and organs **(Pander 1817)**. His work was continued by Karl Ernst Von Baer, who expanded the germ layer concept to all vertebrates **(Von Baer 1828)**. This discovery, together with the progressive development of the cell theory in the mid-19th century, clearly favored epigenesis, and sounded the death knell of preformationism –at least in its old, literal form.

But despite careful observation of developing embryos, epigenesis lacked a convincing causal mechanism for the organization of the initial mass of cells into an adult animal. In the absence of a robust theory of embryonic organization, many natural philosophers assumed the existence of a vital force, a life-specific *vitalis* responsible for the organization and complexification of the embryo throughout its development (for a short review, see **Betchel and Richardson 1998**). Such a vital force conveniently filled the theoretical gap of epigenesis, but also relied on the idea that life was irreducible to purely physical and chemical laws, and in that respect remained compatible with spirituality. However by the end of the 19th century, and later in the early 20th century, a vast majority of biologists would progressively abandon vitalism, and rather seek for deterministic, testable mechanisms of embryogenesis. Since then, a major axis of biology has become to identify these mechanisms by studying the behavior of cells and tissues, and more recently, of genes. Indeed, in the second half of the 20th century it was shown that our genes, by controlling cell behavior and eventually cell fate, could regulate development and dramatically affect embryogenesis. Today, genetics has, without a doubt, become the prevalent framework to study the mechanisms of embryogenesis.

Yet at the beginning of the 20th century, a few attempts at providing a theoretical framework to epigenesis were made. Following Lamarck's idea that life should stem from physical and chemical principles only, German chemist Moritz Traube initiated a series of attempts at creating "artificial cells" at the end of the 19th century **(Traube 1866)**. His works influenced many, notably the chemists Denis Monnier and Carl Vogt, with their work on the artificial production of organic forms **(Monnier and Vogt 1882)**. About twenty years later, at the beginning of the 20th century, French physician Stéphane Leduc turned this emergent field of investigation into a proposal for a paradigm shift in biology. Not only he defended that life was a purely physical phenomenon, and that no life-specific principle, no vital force, should exist; but he also proposed that both apparition of life and embryogenesis should result from the intrinsic power of organization of physical forces, with the idea that physical interactions should be able to organize matter in a spontaneous fashion, in living organisms just as in the inert world. And to support his theory, Leduc took the artificial synthesis of life-like elements one step further, and developed synthetic experiments from purely physical ingredients, in which he mimicked a variety of shapes and phenomena only observed in living organisms: cellularization, cell division and karyokinesis, plant-like growth, seashell-like crystallization, etc.

Touching too closely to the question of spontaneous generation, Leduc's work was excluded from the reports of the French Académie des Sciences, and was largely forgotten later in the 20th century. Most of the credit of introducing a physical approach to biological morphogenesis rather went to D'Arcy Thompson, although he made extensive use Leduc's work in *On Growth and Form* **(Thompson 1917, p. 293, p.297, p.324, p.500-503, p.660, p.664, p.854)**. At best, Leduc was remembered for his amusing bio-mimetic experiments, or for having first used the term "synthetic biology". Yet his approach makes a great deal of sense in regard of modern physics and of the most recent advances of developmental biology. The physics of morphogenesis has greatly developed in the second half of the 20th century, and elucidated many mechanisms of spontaneous organization in solids, fluids, sand, and other so-called complex systems; showing that simple rules of interactions between the parts of a system could spontaneously generate complex, organized patterns in a robust manner. These discoveries, which led to the modern concepts of emergence and self-organization, are more and more applied to living systems. And today, developmental biology, the modern science of embryogenesis, increasingly turns to physics and self-organization to identify mechanisms of organization and morphogenesis during embryonic development.

In this article I will give an overview of Leduc's theoretical work, and of his related experiments. I will analyze the proximity of his ideas to the concepts of emergence and self-organization, which today begin to pervade the field of developmental biology. I will discuss the paradoxical oblivion of physical approaches such as Leduc's to problems of living organization in the 20th century, and in particular during the genetics era. I suggest that our reluctance to abandon the "vital exception" has played a major role in this process, and that recent developments of embryogenesis might prefigure a paradigm shift, as advocated by Leduc at the beginning of the 20th century.

**Overview of Leduc's work**

Stéphane Armand Nicolas Leduc (1853-1939, **Figure 1**) was first trained in the physical sciences, and then defended a thesis of medicine in 1883. He became a general practitioner and was granted a chair of medical physics in the city of Nantes, France. As a doctor in the 1880's, he dealt with the epidemics cholera and typhoid, and worked on public health issues such as clean water and sewage systems. In parallel, he carried out medical research in various fields, from electrophysiology to radiotherapy **(Drouin et al. 2014)**. Only later, at the onset of the 20th century, did he develop his new ideas about biological organization, in the essentially nonexistent context of biophysics. Most notably, and in addition to numerous articles and conferences published before the exclusion of his work from the French academy of sciences, he published three books related to embryogenesis and biological organization: *Théorie Physico-chimique de la Vie et Générations Spontanées* **(Leduc, 1910)**, revised and re-edited in english as *The Mechanism of Life* **(Leduc 1911)**, *La Biologie Synthétique* **(Leduc 1912)**, and later *Enérgétique de la Vie* **(Leduc 1921)**. In these books Leduc exposed in details his physical theory of life, a few years before D'Arcy Thompson published the acclaimed *On Growth and Form* **(Thompson 1917)**. The theoretical side of Leduc's work deals with epistemological considerations on the physical nature of life, and the practical consequences it should have on the study of life phenomena. In particular Leduc deplores the "mysticism" of his colleagues and advocates the use of physical methods in biology. The experimental side of his work consists in bio-mimetic experiments that illustrate his physical theory of life. He introduces "synthetic biology," the synthesis by osmosis and diffusion of shapes strikingly similar to that of living systems, and argues that such physical mechanisms might be good candidates for both morphogenesis –the organization of the embryo during its development– and biogenesis –the emergence of life.

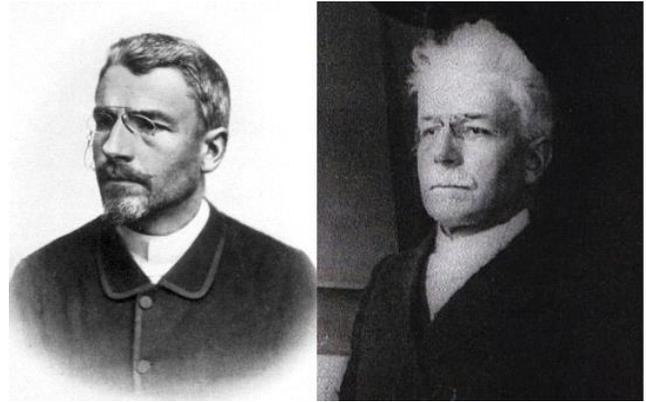

**Figure 1 –** Stéphane Leduc (1853-1939)

**Theory -** Leduc's obsession in adopting a new physical paradigm clearly demonstrates that he positioned himself in a theoretical frame of biology. In his *Synthetic Biology*, he insists on the scientific necessity of having theoretical guidelines, particularly in biology: "The absence of a general theory of life greatly harms the progress of biology" **(Leduc 1912, chapter 2)**. Like Ernst Haeckel, whom he greatly admires, Leduc thinks that Lamarck deserves a place beside Darwin in the history of evolution **(Leduc 1911, p.164, citing Haeckel)**, and it is in particular the theoretical motivations of Lamarck that strike Leduc: "It is with the idea that nothing is more hindering than the absence of a general theory that Lamarck edified his theory of evolution" **(Leduc 1912, chapter 2)**. Indeed, Lamarck introduced the first theory of evolution as a theoretical necessity, the spontaneous generation of higher organisms being impossible, it was required that simpler organisms could evolve, diversify and complexify over generations. Lamarck also tried to explain how they could do so in his transformist theory **(Lamarck 1809, part 1 - chapter 7)**, which as we know today was proven inexact. Fifty years later, the theory of evolution was corrected by Charles Darwin, who introduced natural selection of characters as the central mechanism for the evolution of species **(Darwin 1859)**. Yet it is the theoretical achievement by Lamarck of even postulating evolution that Leduc admires, more than Darwin's answer to how they do so: "If the transformist theory of evolution is abandoned, it yet remains a powerful agent of progress […]; and its authors, pioneers and heroes" **(Leduc 1912, *chapter 2*)**. Leduc's work on morphogenesis is in essence dedicated to fill the theoretical gap of epigenesis, gap largely left unfilled by Darwin's theory of evolution. Darwin was indeed surprisingly reluctant to discuss the origin of life, as pointed out by Ernst Haeckel: "The chief defect of the Darwinian theory is that it throws no light on the origin of the primitive organism –probably a simple cell– from which all the others have descended. When Darwin assumes a special creative act for this first species, he is not consistent, and, I think, not quite sincere" **(Haeckel 1962)**. Leduc thus addresses the question of the origin of life from inorganic ingredients, which in his mind should be intimately related to the mechanisms underlying the organization of the embryo: "The Darwinian theory shows how acquired variations are transmitted and accentuated by natural selection, but it says nothing as to how these variations may be acquired. In the same way we are in entire ignorance as to the physical mechanism of ontogenetic development, the evolution of the embryo" **(Leduc 1911, p.167)**.

**Experiments -** Like several theoretical biologists of his time (Edward Schäfer, Jacques Loeb, Alfonso Herrera, to name a few), Stéphane Leduc defended a mechanistic conception of life, and therefore advocated the use of physical methods in the life sciences. He vividly and repeatedly insisted that "physicism" should replace "mysticism" in biology: "Two methods exist to explain nature's phenomena: mysticism and physicism. Physicism is the method of the physical sciences, mysticism still prevails in biology" **(Leduc 1912, chapter 1)**. With that said, Leduc attempted to make biology synthetic: "Until now, biology relied on observation and analysis only. The use of observation and analysis, without a synthetic methodology, is one of the causes that delay the progress of biology. The analytic method in biology is paralyzed, sterilized, by the indissoluble unity of phenomena: if, in a living system, one tries to isolate a phenomenon from the others, the phenomenon disappears, the animal dies. Not only the synthetic method is applicable to biology as it is for other sciences, but it seems to be the most fertile, the most capable to reveal the physical mechanisms of life, the study of which has not yet begun. When a phenomenon, in a living organism, has been observed, and that one believes to understand its physical mechanism, one should be able to reproduce this phenomenon separately, outside of the living organism" (Leduc 1912, chapter 2). Leduc thus designed experiments in order to mimic behaviors and shapes found in living systems from purely physical and chemical ingredients.

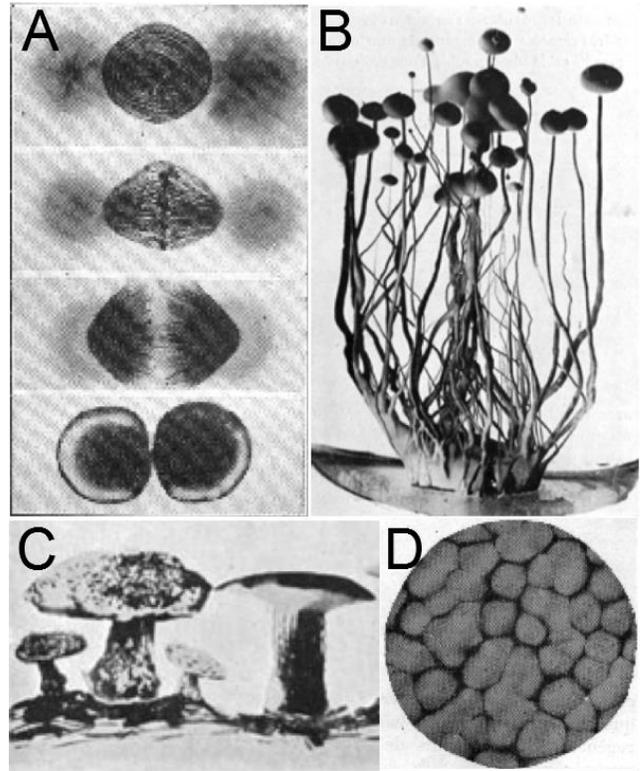

**Figure 2** – Growth figures **(Leduc 1912)**. **A.** Artificial karyokinesis obtained by diffusion in a saline solution. **B.** Artificial plant-like osmotic growth. **C.** Artificial osmotic shells. **D.** Artificial segmentation in a solution of nitrate potassium.

Among several other techniques, Leduc extensively utilized principles of diffusion and osmosis, to show that shape emergence as seen in the living world was indeed within the reach of the physical and chemical forces alone. Most of Leduc's work on morphogenesis is based on the spontaneous osmotic growth of salts of heavy metals in liquid solutions such as sodium silicate (or liquid glass). Chemical precipitation occurs at the interface between the salt and the liquid glass, forming a semi-permeable membranes, an artificial "cell" enclosing the crystals. Then osmotic pressure tends to equilibrate the salt concentration on both sides of the semi-permeable membrane, which results in water flowing inside the cavity. At some point, pressure exerted by water inside the cavity breaks the membrane, and the salt trapped inside is released in the surrounding solution for a short period of time, until a new membrane forms by precipitation. The repetition of this process causes the crystal to slowly grow in the solution. Depending on the type of crystal, the concentration of the solutions, and the concentration gradients used to prepare the water glass medium, a great variety of patterns and growth types can be obtained, which, Leduc showed, can be strikingly similar to simple life forms or phenomena **(Figure 2)**.

Although the physical and chemical mechanisms underlying osmotic growth are now well understood **(Cartwright et al. 2002, Barge et al. 2015)**, Leduc's experimental skill in creating such shapes remains unmatched. The spontaneous organization of Leduc osmotic creations into life-like forms such as plants, fungi, shells, or cells, and into life-like events such as branching, karyokinesis, or cellular segmentation, supported his view that physical forces alone are sufficient to organize living matter. His spectacular experimental achievements, as compared to other theorists advocating synthetic biology, such as Loeb or Schäfer, definitely contributed to build Leduc's influence in the early 1900s. Beyond shape similarities, Leduc attributed a variety of physiologic functions, such as nutrition (transformation of an external source of energy into organized shape changes), growth, or reproduction, to his osmotic structures: "Evolution, nutrition, sensibility, growth, organization, none of these, not even the faculty of reproduction is the appanage of life" **(Leduc 1911, p.xiii)**. And beyond embryogenesis, Leduc proposed that the chemical criteria for such osmotic growth might have been met in the early ages of earth, and that such processes might have driven the spontaneous appearance of life on earth: "Is it possible to doubt that the simple conditions which produce an osmotic growth have frequently been realized during the past ages of the earth?" **(Leduc 1911 p.144)**.

**Leduc, vitalism, and spontaneous generation**

Leduc's theory was questioning the very nature of life. It is one thing to accept that life should stem from physical and chemical principles, but Leduc took that belief a step further, with the idea that none of these principles should be specific to life. While he candidly asked the question as to whether these osmotic forms might be considered as alive, his detractors mocked the pale imitation that they were, evocating for instance "a resemblance as superficial as that between a person and his marble image" **(Oparin 1938, p.56)**. Yet, Leduc's question was not that naïve. Until now no definition of life can be given, since defining life requires a purely arbitrary line between living and inert, on a basis which is actually more ideological than scientific. Many characters that we intuitively associate with life, such as growth, death, transformation of an external income of energy into shape changes, homeostasis, or replication are indeed not specific to life, and can be found in systems widely considered as inorganic. The transition between inorganic, inert objects and living, organic objects is rather continuous than sudden, and therefore difficult, if not impossible, to characterize. This was coined by Leduc in perfectly simple terms: "Since we cannot distinguish the line between life and the rest of nature's phenomena, we should conclude that this line does not exist, which satisfies the law of continuity between all phenomena" **(Leduc 1910, p.14)**. "Life is difficult to define because it differs from one living being to another; the life of a man is not that of a polyp or of a plant, and if we find it impossible to discover the line which separates life from the other phenomena of Nature, it is in fact because no such line of demarcation exists. The passage from animate to inanimate is gradual and insensible" **(Leduc 1911, p.xiii)**. Thus, one can understand Leduc's point when he asked whether osmotic structures displaying growth (and growth arrest), segmentation, or internal fluid circulation might be called "alive".

And trying to narrow or even suppress the gap between animate and inanimate, Leduc met fierce vitalist resistance, or at least a reluctance to abandon the vital exception. The epigenetic theory of embryogenesis lacking a theoretical basis, the idea that a vital force could support the organization of embryos had long been a most convenient one; in fact since Caspar Friedriech Wolff's, considered as one of the founders of modern embryology, introduced the concept of a formative vital force he called *vis essentialis* **(Wolff 1759)**. The scientific role of the vital force in the late 18[th] and in the 19[th] century, can in some respects be compared to that of ether, the versatile physical entity supposed to support gravity or light propagation in vacuum. In short, a convenient medium accounting for interactions beyond our understanding, and possibly tinged with ideology and spirituality. Interestingly, Wolff declared that "all believers of epigenesis are vitalists". Since epigenesis, unlike preformation, implied that God had not preformed all animals, vitalism was also a way to rescue epigenesis from miscreance. And while vitalism no longer was the dominant doctrine a hundred years later, vitalism was still very much present in the life sciences, at least in an insidious form. According to Leduc: "Mankind increasingly replaces conjuration to gods by efforts of reason. Almost alone, biologists and physicians still conjure up mysterious forces. Old anthropomorphism, finalism, supernatural, metaphysical, extra and ultra-scientific persists in the life sciences, under various forms and with various intensity. One admits there principles that are beyond matter, a vital force specific to life, and the finalism is met everywhere. Education in this regard is such that, even the works that claim the physico-chemical nature of life are sprinkled of vitalist interpretations and finalist explanations. The phenomena of life are considered with authentic superstition, it is a sacrilege to try and interpret them, and doing so with the methods of physicism, one rises the most violent oppositions" **(Leduc 1912, chapter 1)**. And indeed vitalism was far from extinct. Louis Pasteur, for instance, considered fermentation to be irreducible to ordinary chemistry, and therefore a vital phenomenon **(Pasteur 1879)**. Hans Driesch, a German embryologist, having observed that isolated pieces of blastomeres could develop into complete embryos, supposed that this developmental persistence was irreducible to mechanistic explanations and invoked an autonomous principle he called entelechy (see for example **the Gifford lectures, Driesch 1908**). French philosopher Henri Bergson, in his *Creative Evolution,* introduced his own concept of "*élan vital*" or *"vital impetus"*; and attacked synthetic approaches such as that of Leduc. Without naming him explicitly, Bergson describes his work in harsh terms **(Bergson 1907, p21-23)**. "Imitation of the living by the unorganized may, however, go a good way. Not only does chemistry make organic syntheses, but we have succeeded in reproducing artificially the external appearance of certain facts of organization, such as indirect cell division and protoplasmic circulation. […] But scientists are far from agreed on the value of explanations of this sort. Chemists have pointed out that even in the organic – not to go so far as the organized– science has reconstructed hitherto nothing but waste products of vital activity". Then Bergson enthusiastically refers to the words of Edmund Beecher Wilson, an American cell biologist: "The study of the cell has, on the whole, seemed to widen rather than narrow the enormous gap that separates even the lowest forms of life from the inorganic world". Leduc denied the existence of such a gap, and on the opposite was very attached to the "law of continuity" between all phenomena. Hence he insisted that life phenomena should not be studied differently than physical phenomena: "Life is a purely physical phenomenon, produced by the same forces, ruled by the same laws that govern the non-living world" **(Leduc, 1912, chapter 2)**. Hence, "the problem of biology is to understand how physical forces act on living matter" **(Leduc, 1912, chapter 3)**.

And to replace the "mysticism" of vitalism, Leduc called upon the inherent power of organization of physical interactions: "The ordinary physical forces have in fact a power of organization infinitely greater than has been hitherto supposed by the boldest imagination" **(Leduc 1911, p.167)**. This concept of physical spontaneous organization, as introduced both theoretically and experimentally by Leduc, is remarkably similar to the later concepts of emergence and self-organization, which proved to be powerful physical frameworks for the understanding of pattern formation, both in the living and non-living worlds. Importantly, Leduc's theory of living self-organization is much distinct from pre-existing principles proposed by earlier epigenists, such as Caspar Friedrich Wolff's *vis essentialis* ("the very power through which, in the vegetable body, all those things which we describe as life are effected", see **Gigante 2009, p.19**) or Johann Friedrich Blumenbach's *bildungstrieb* (powers that "are not referrible to any qualities merely physical, physical or mechanical", see **Gigante 2009, p.16**). Although these are enunciated as autonomous principles, they rather remain avatars of a vital force. Emmeche et al. wrote that "there is a very important difference between the vitalists and the emergentists: the vitalist's creative forces were relevant only in organic substances, not in inorganic matter. Emergence hence is creation of new properties regardless of the substance involved". Later they conclude that "the assumption of an extra-physical vitalis (vital force, entelechy, élan vital, etc.), as formulated in most forms (old or new) of vitalism, is usually without any genuine explanatory power" **(Emmeche et al.1997)**. Clearly Leduc's concept of physical organization is much closer to emergentism (or self-organization), and his experiments were designed to prove that indeed no specific force is required to organize living systems, since the ordinary physical forces possess this very power of organization. And as naïve or over-interpreted as they are, his experiments still proved that physical interactions could spontaneously organize non-trivial shape changes, in a manner that can at least *resemble* life, and even "deceive the very elect", as written by Deane Butcher in his preface to Leduc's *Mechanism of Life* **(Leduc 1911, p.viii)**: "There is, I think, no more wonderful and illuminating spectacle than that of an osmotic growth,—a crude lump of brute inanimate matter germinating before our very eyes, putting forth bud and stem and root and branch and leaf and fruit, with no stimulus from germ or seed, without even the presence of organic matter". For Leduc, this was a proof of principle that the laws of physics should be sufficient to organize matter into life, both for the development of a single organism, and for the appearance of life on earth. With a certain logic, Leduc called this self-organization "spontaneous generation", and argued that such spontaneous generation is required in the theory of evolution: "The chain of life is of necessity a continuous one, from the mineral at one end to the most complicated organism at the other. We cannot allow that it is broken at any point, or that there is a link missing between animate and inanimate nature. Hence the theory of evolution necessarily admits the physico-chemical nature of life and the fact of spontaneous generation" **(Leduc 1911, p.15)**. In short, "without the idea of spontaneous generation and a physical theory of life, the doctrine of evolution is a mutilated hypothesis without unity or cohesion" **(Leduc 1911, p.164)**. However, Pasteur's experiments in sealed bottles had just ended the academic debate that set he and Pouchet in vivid opposition about the spontaneous generation of life from inorganic materials, showing that such generation could not occur if proper precautions were taken. And yet Leduc insisted that the issue of spontaneous generation had not been settled, for it eventually was that of the origin of life: "The question of spontaneous generation exists, and it's not in the power of anyone to suppress it. It is stupefying that Pasteur's experiments could extinguish it so completely for more than thirty years" **(Leduc 1912, chapter 15)**. As discussed above, what Leduc called spontaneous generation was largely different from the spontaneous generation defended by Pouchet as dismissed by Pasteur, and rather akin to an autonomous physical organization. Even though it precisely covered the concept Leduc wanted to designate, the use of the terminology of "spontaneous generation", proved to be a regrettable choice. As Dean Butcher mentioned in his preface to *The Mechanism of Life*, to explain the reasons of Leduc's exclusion from the reports of French Académie des Sciences, his writings "touched too closely on the burning question of spontaneous generation". Leduc himself wrote that "in 1907 the Académie des Sciences de Paris excluded from its *Comptes Rendus* the report of my researches on diffusion and osmosis, because it raised the question of spontaneous generation". One might argue that Leduc, being aware of the controversy, should have insisted less on the concept of spontaneous generation. But the global dismiss of spontaneous generation was precisely a persistence of the vitalism he fought, exactly because in his opinion a theory of spontaneous generation was required to get past the vital exception. This recklessness might also indicate yet another reason for the rejection of Leduc's work. One can sense from his bold writing style that he did like to shock the reader and oppose the mainstream. On several occasions he mocks the methods or beliefs of his colleagues in rather harsh words. I suggest this bears witness to his tendency to consider himself as a misunderstood avant-gardist. Clearly, Leduc had a special admiration for men whose ideas were so innovative that they were rejected by their contemporaries. On multiple occasions, he cites Galilée, Christophe Colomb, Giordano Bruno, or other such men, and there is little doubt that he likes to consider himself as one of them: "The explorer of the unknown must be aware that, leaving the most used routes, where rich cities and comfortable areas can be found […], he will find solitude. If there he meets other minds, they can be nothing but an elite, and would cease to be an elite if they became the majority […]. Ignored great men, in their understanding of what is universally misunderstood, have a motive a supreme pride" **(Leduc 1912, chapter 16)**. One can only hypothesize that such an attitude, coupled to his extreme disregard of traditional methods used in biology (which he refers to as mysticism!), might have facilitated the rejection of his already marginal work by his peers.

And indeed, Leduc is almost absent from modern scientific literature, even though his ideas were quite debated in his time, in France and abroad. In comparison with France, Leduc's reception in the English and American scientific press, while mitigated, turned out to be more sympathetic (see Keller's detailed analysis, **Keller 2002, p.29-35**). If some commenters were dubious about some of Leduc's conclusions, his assertion that vital phenomena should originate from basic physical principles "tapped a nerve that was very much alive in the English speaking world," as coined by Keller. Leduc, while often mocking the "frozen" erudition of French scientists, had great consideration for the "country of liberty" (England) and for the Anglo-American people, a fertile ground for "the birth and development of new ideas" **(Leduc 1912, p.ix)**. It is then no surprise that Leduc, with the help of Deane Butcher, published his *Mechanism of Life* in English in 1912. Yet, abroad just as in France, Leduc's work was soon forgotten and lost most of its influence. As suggested by Evelyn Fox Keller, the threat of vitalism had almost disappeared, and the subsequent necessity for a physical theory of life had become accordingly less urgent **(Keller 2002, p.48)**. After the advent of genetics, and during most of the 20$^{th}$ century, physical approaches to embryogenesis almost disappeared from the life sciences. Yet, the idea that physical interactions possess a power of organization susceptible to organize living matter has made its way to the life sciences and echoes in the most recent advances of developmental biology. In the following, I will first discuss the decline of physical approaches inherited from Leduc or D'Arcy Thompson during the 20$^{th}$ century's genetics era; then the recent developments of developmental biology that are leading to what, I suggest, might be called their rehabilitation.

**Leduc and modern biology**

The development of genetics since the discovery of DNA structure in 1953 **(Watson and Crick 1953)** transformed our conception of life. It also provided new and powerful tools for the study of the mechanisms of life and of embryogenesis. With the astonishing discovery of a molecular basis underlying the regulation of embryonic development, genes were soon considered the elementary blocks of life; and with the technological development of molecular tools, biology soon became focused on the decryption of the genetic code and its regulation of cellular processes. This shift towards a genetic view of life is first exemplified by the methods now used in biology, almost exclusively turned to molecular and genetic tools. Second and equally important, the gene has become a cultural icon of life and everything related to it: disease of course, but also behavior, personhood, identity, skill, etc. The gene, and the genetic determinism associated to it, are now part of the popular culture, to be found in novels, movies, cartoons, or video games. And this process has been continuously fed by science: every week or so we hear of a new gene that has been shown to affect autism, depression, obesity, sexuality. The widespread use of –originally– metaphoric expressions, such as the "genetic program" our genes assemble into, or that our genome is the "book of life", bears witness to the deep anchorage of the gene mystique into the collective imagination. It has now been a century or so that biology has rejected vitalist theories, and yet it is striking how vitalistic these expressions sound. Clearly the rapid spread of the genetic terminology has to do with its simplicity and its specificity to life phenomena: in the absence of a defining feature of life, we naturally welcomed genes as the underlying basis of life-specific traits, basis that could carry all the information specific to an individual, including a human individual. As coined by Jean-Jacques Kupiec, geneticist and epistemologist **(Kupiec 2002, interview to Ecorev')**: "Biology has not made its Copernican revolution yet"; in the sense that the vital exception, which underlies our own human specificity, remains at the center of biology. The scientific and popular success of the gene mystique bears witness to our reluctance to give up on this specificity, and accordingly to our eagerness to adopt scientific and cultural frameworks that preserve it. As Kupiec said, "Genetics is not a scientific theory anymore, it's an indeology" **(Kupiec 2002)**. This unrestrained scientific and popular geneticization of life can be considered, I suggest, as a new form of vitalism, in the sense that genes replaced the vital force that bestowed life its precious specificity. This is perfectly exemplified by James Watson's formula: "We used to think our fate was in our stars; now we know, in large measure, our fate is in our genes".

In that context it is not that surprising that genetics eclipsed most of the other approaches to life sciences, *a fortiori* the fundamentally unspecific physical approaches. Yet this came to a price. While succeeding in identifying causal cascades (e.g. without this or that gene, a modification of the phenotype is observed), a purely genetic approach often fails to identify mechanisms that link gene expression to physical traits. Genome sequencing, considered at the time of the Human Genome Project as a huge step towards our understanding of life phenomena, has left a somewhat bitter taste to the scientific community. If anything, it showed that the genome is *not* the book of life. The editorial of Nature's special "Human genome at ten", published ten years after the sequencing **(Nature editorial, 2011)**, bore the self-explanatory title "Best is yet to come". The distance between promises and achievements reflected the fact that "the project was a triumph of technological capability rather than scientific understanding" **(Ball 2010)**. Philipp Ball, editor for the Nature journal and scientific writer, wrote in an article written for Prospect Magazine: "[An] unfortunate notion behind the Human Genome Project was that science can be done without hypotheses or ideas." Ball then quotes Jim Collins of Boston University, whom he refers to as "one of the few biologists to see a bigger picture": "We've made the mistake of equating the gathering of information with a corresponding increase in insight and understanding" **(Ball 2010)**. And indeed, the massive amount of data provided by the Human Genome Project is hard to exploit without a more general theoretical framework.

Hence Sonigo, Kupiec, and others, predict that the difficulties met by the genetic paradigm prefigure a paradigm shift, the "Copernican revolution" mentioned earlier. Very much like the shift proposed by Stéphane Leduc, this revolution requires a more general theoretical framework, unbiased by the specificity of life. Several clues indicate that we might be already witnessing the forerunners of this shift. First, quantitative tools, physical methods, and modelling – the "physicism" advocated by Leduc– have literally invaded biology over the last few years. Second, and maybe more importantly, biology increasingly turns to general concepts of physical self-organization to decipher actual mechanisms of emergence, integrating to our understanding of embryogenesis the "power of organization of physical forces" coined by Leduc.

In the field of embryogenesis, several gaps in the purely genetic paradigm might have facilitated its loss of impetus, and conversely the advent of more general physical concepts. Genetic patterning of the embryo, as crucial as it is, cannot alone provide actual mechanisms of shape formation without physical considerations. How can shapes be generated by homogeneous or non-homogeneous growth? How is shape generation impacted by the mechanical nature of living tissues? Is there any crosstalk between mechanical state and genetic regulation? How do shape boundaries, by limiting diffusion, feedback on genetic patterning, which relies on the physical diffusion of proteins? The global description of the feedback between shape, gene regulation, and growth requires a physical backdrop; and eventually the framework of embryonic development has to be the physics of its constitutive elements, such as cells and tissues. Even though interactions between these elements can be mediated by genes, they have to be translated into physical units of growth, shrinkage, motion, flows, etc. Equally important, shapes can appear in biological systems independent of genetic regulation, because physical interactions are morphogenetic in essence. Everywhere inorganic, *geneless* systems demonstrate the morphogenetic power inherent to the physical laws of interactions. There is of course no reason that such self-organized morphogenesis should be absent from life phenomena, and in particular it should be of crucial importance in embryonic development. This idea, which is remarkably close to Leduc's theory of spontaneous generation, only begins to pervade the field of developmental biology. While one might consider that Leduc's experimental data was insufficient to support such strong theoretical claims, his new physical approach was, as I will discuss now, consistent with the most recent and groundbreaking advances of developmental biology. In the following, I will provide three examples, meaningful because of their general character, among the many and growing number to choose from: the gastrulation of embryos, the formation of vertebrates' body plan, and plants' phyllotactic organization.

**Gastrulation -** Gastrulation is an early phase of the embryonic development common to most animals. It is the archetypal process by which the blastula, a primitive hollow sphere of cells, folds into a layered structure to form the endoderm, the ectoderm, and the mesoderm, from which all internal organs will be formed. As amusingly coined by Lewis Wolpert, "It is not birth, marriage, or death, but gastrulation, which is truly the most important time in your life." Recently, it was shown in *Drosophila* by Eric F. Wieschaus and collaborators that gastrulation highly relies on the hydrodynamic nature of the embryo, which transmits apical constrictions exerted by the cellular cytoskeleton to the rest of the tissue, and causes the passive inward flow of cells called invagination that eventually forms the mesoderm **(He et al. 2014)**. This is in contrast with previous interpretations stating that genetically coordinated active cell shape changes should cause invagination. And indeed, additional experiments by the Wieschaus group showed that mutant individuals that do not form cells during the so-called cellularization process still undergo invagination. Interestingly, the apical constriction that triggers the flow relies on genetic expression. Whether any crosstalk exists between the genetic activity that triggers constriction and the mechanical state (that is, tension, compression, or any deviation from the reference state) of the cells remains subject to debate, yet several papers point out that genes can be mechanosensitive (for a short review, see **Jaalouk and Lammerdink 2009**), in particular in this very system **(Farge 2003, Brunet et al. 2013)**. Anecdotally, the fact that Wieschaus was awarded the Nobel Prize in 1995 for his earlier work on the control of embryogenesis by genetic signaling pathways reveals how much the spirit has recently changed in developmental biology.

**Body plan -** The hydrodynamic, or more precisely, viscoelastic nature of cell and tissue movements has proven to be a powerful tool to study shape emergence in early embryogenesis. It was also utilized by Vincent Fleury and others as a framework to study the formation of the vertebrates' body plan. The term "body plan" encompasses basic morphological features common to the animals of a phylum. Typically, these features can be symmetries, limb disposition, or segmentation of the main axis into repetitive segments (such as the vertebral column). It is largely unknown if the establishment of the body plan results from a nonlinear succession of "stop-and-go" genetic instructions, with no global routine, or if it somehow follows an "archetype", as coined by Darwin himself. Fleury showed that basic viscoelasticity and mass conservation laws applied to the early embryo yield an hyperbolic symmetry breaking, that can be observed in the form of highly conserved hyperbolic tissue flows, and that results in the axial elongation of the embryo and in the four-limb symmetry observed in most vertebrates **(Fleury 2012)**. This suggests that archetypes might indeed exist, as set by the formative laws of physics. Importantly, this should not be mistaken for the *a posteriori* selection of shapes that fit well the external physical constraints. It is rather the *a priori* restriction of developmental shapes to those made accessible by the laws of physics: the space of shapes that are actually accessible to development is generated by the laws of physics, and consequently physical interactions can be said to be formative and to define the archetypes.

**Phyllotaxis -** Phyllotaxis is the striking geometrical arrangement of plant organs along the stem. It has been observed centuries ago that leaves organize on the plant stem into spectacular spirals described by the famous mathematical suite of Fibonacci. Yet the mechanism by which they could do so has long challenged our understanding of plant development. German botanist Wilhelm Hoffmeister was the first to propose an organization rule that he believed might be sufficient to generate phyllotactic spirals: he proposed that primordia (new organ precursors) should periodically form at the growing tip *where there is the most space available* **(Hoffmeister 1868)**. It took until the end of the 20$^{th}$ century and the work of Stéphane Douady and Yves Couder to refine Hoffmeister's theory and to show that a simple inhibition rule between successive primordia (the appearance of new primordia being inhibited in the region of newly formed ones) was mathematically sufficient to spontaneously generate the Fibonacci spirals **(Douady and Couder 1992)**. Using this finding, they could mimic plant phyllotaxis in a purely physical experiment, much in Leduc's spirit: successive drops of ferrofluid (mimicking the primordia) fall on a dish and are advected by a magnetic field generating repulsive forces between drops. And indeed the drops spontaneously arrange into Fibonacci spirals. They later showed that a variety of observed phyllotactic patterns were in fact different regimes of this simple mechanism **(Douady and Couder 1996)**. The transport and accumulation of the hormone Auxin near newly formed primordia was later found to underlie growth inhibition **(Reinhart 2003)**. Since then, the feedback between hormonal cues, phyllotactic patterns and tissue mechanics at the shoot meristem (the growing tip of the plant) has become an active field of investigation in plant development. Phyllotaxis nicely illustrates how simple interaction rules can force the spontaneous convergence of a system towards archetypal shapes, independent of genetic regulation or selection.

## Conclusion

Stéphane Leduc's legacy is not limited, I suggest, to his osmotic experiments, but rather encompasses his whole physical approach to the problem of organization in biology. Evelyne Fox Keller exhumed Leduc's work to illustrate the historical role of physics in understanding the mechanisms of life **(Keller 2002)**. Like Keller, I stress that Leduc's theory has a significant importance in the history of our attempts at understanding life phenomena. Yet Keller, like Pierre Thuillier before her **(Thuillier 1980)**, suggested that Leduc's works are "illuminating in proportion to what may now appear to us as their absurdity" **(Keller 2002, p.24)**. If indeed some of his claims might now seem outdated, he introduced the idea and some experimental evidence that the ordinary physical forces might intrinsically possess a power of organization susceptible to organize living matter during embryonic development. This concept was remarkably ahead of its time and, as I discussed, is akin to today's bio-physical self-organization.

Yet, Leduc's theoretical contribution to biology fell into oblivion for a century or so. I suggest that this is mostly a consequence of our reluctance to adopt concepts and tools that are unspecific to life, because they threaten the specificity of living things –the vital exception; and therefore our own specificity. Just like Copernic when he introduced his heliocentric model instead of the commonly accepted geocentric model, Leduc and others had to fight an ideological resistance. The 20$^{th}$ century, the "century of the gene" as Evelyn Fox Keller has called it, has established genetics, a truly life-specific approach, as the framework of the life sciences. Remarkably easily and rapidly, the gene and the DNA strand have become popular icons of life and of our biological identity. Yet, the end of the 20$^{th}$ and the beginning of the 21$^{st}$ century have demonstrated the difficulty for the genetic paradigm to identify explicit mechanisms relating genetic information to physical traits. And most recent developments of biology, in particular in the field of embryogenesis, show that physical concepts inherited from Leduc or D'Arcy Thompson can help identifying these mechanisms. This conceptual change of spirit in modern biology might be a forerunner of a paradigm shift initiated more than a century ago.

Finally, Leduc's story bears witness to a rather paradoxical aspect of 19$^{th}$ and 20$^{th}$ century's biology: How can biology be the science of life phenomena while avoiding the pitfall of the vital exception? A hundred years ago, Leduc stressed that physical methods should prevail over life-specific methods. The modern revival of physical approaches, often used beside now conventional genetic tools, might be rooted in the same need to address this paradox.